\documentclass[9pt,twocolumn,twoside]{osajnl}

\journal{ol} 

\setboolean{shortarticle}{true} 

\title{Passively synchronized dual-color mode-locked fiber lasers based on nonlinear amplifying loop mirrors}

\author[1]{Jing Zeng}
\author[1]{Bowen Li}
\author[1]{Qiang Hao}
\author[2]{Ming Yan}
\author[1,*]{Kun Huang}
\author[1,2,3,4]{Heping Zeng}

\affil[1]{Shanghai Key Laboratory of Modern Optical System, and Engineering Research Center of Optical Instrument and System, Ministry of Education, School of Optical Electrical and Computer Engineering, University of Shanghai for Science and Technology, Shanghai 200093, China}
\affil[2]{State Key Laboratory of Precision Spectroscopy, East China Normal University, Shanghai 200062, China}
\affil[3]{Jinan Institute of Quantum Technology, Jinan, Shandong 250101, China}
\affil[4]{e-mail: hpzeng@phy.ecnu.edu.cn}
\affil[*]{khuang@lps.ecnu.edu.cn}

\dates{Compiled \today}

\ociscodes{(060.3510) Lasers, fiber; (140.4050) Mode-locked lasers; (190.4370) Nonlinear optics, fibers.}

\doi{\url{http://dx.doi.org/10.1364/XX.XX.XXXXXX}}

\begin{abstract}
We have proposed and implemented a novel scheme for passive all-optical synchronization between erbium and ytterbium mode-locked fiber lasers. The passive locking of repetition rates for the dual-color pulses was realized by cross-phase modulation within phase-biased nonlinear amplifying loop mirrors. In contrast to previous demonstrations, the synchronization system was configured in an all-polarization-maintaining structure, thus gaining substantially improved stability and robustness. Consequently, the maximum tolerance of cavity-length mismatch of 16.2 mm was achieved unprecedentedly, which was at least one order of magnitude longer than previously reported results for comparable temporal durations of involved pulses. The corresponding relative timing jitter was measured to be 31 fs within 1-MHz bandwidth. Such tight and robust synchronization fiber laser system offers a great potential for various applications, such as pump-probe microscopy, Raman scattering spectroscopy and nonlinear frequency generation.
\end{abstract}

\setboolean{displaycopyright}{true}

\begin{document}

\maketitle

Spectro-temporal engineering of ultrafast pulses is a prerequisite to develop advanced techniques, typically requiring precise timing of pulse trains and accurate positioning of spectral modes, in modern laser science and technology \cite{Cundiff2003}. In particular, tight and stable synchronization of optical pulses at disparate wavelengths is demanded in various potential applications, such as time-resolved imaging and microscopy \cite{Fischer2016}, Raman scattering spectroscopy \cite{Freudiger2014} and mid-infrared generation \cite{Murray2016}. Although the synchronized dual-color light sources could been accessed based on supercontinuum generation \cite{Dudley2006} or Raman-induced soliton self-frequency shift \cite{Sobon2017}, the available power at specified wavelength windows was usually limited by the conversion efficiency. Moreover, the competition of various nonlinear processes involved in the nonlinear spectral extension would inevitably lead to detrimental intensity instability or diminished optical coherence. Alternatively, synchronized two-color pulses could be obtained from two independent mode-locked laser sources as long as the relative repetition rate is locked \cite{Wei2012, Tian2017}. 

Seminal effort was dedicated in using active techniques like phase locked loops and balanced optical cross correlator \cite{Shelton2002, Miura2002}, which ultimately approached an attosecond relative timing jitter \cite{Kim2010}. Over the past decade, passive all-optical synchronization based on nonlinear cross-phase modulation (XPM) has attracted increasing attention, which enabled the simultaneous realization of simple operation, low cost, and compact layout \cite{Wei2002}. In this case, the needed high-speed feedback to obtain a low timing jitter was provided by the nearly instantaneous response of the nonlinear effect, thus avoiding the stringent requirement of high-bandwidth electronics and complicated feedback system in the active configuration \cite{Zhu2005}. Generally, the underlying synchronization mechanism relies on the combined effect of XPM and group-velocity dispersion, \textit{i.e.} the cavity-length mismatch is compensated by self-adapted group velocity due to the induced spectral shift \cite{Wei2002}.  

Notably, the XPM-based technique has been widely adopted in fiber lasers to implement the passive synchronization \cite{Rusu2004, Yoshitomi2006, Huang2012, Tsai2013}. Indeed, the nonlinear interaction between the involved pulses could be significantly enhanced within the confined and elongated transverse mode of the fiber waveguide. So far, synchronized fiber lasers have been implemented in various passive mode-locking configurations, \textit{e.g.}, based on nonlinear polarization rotation (NPR) \cite{Yoshitomi2006, Huang2012, Tsai2013} or saturable absorber (SA) \cite{Rusu2004, Zhang2011, Sotor2014}. However, all reported synchronization systems were built in a non-polarization-maintaining structure, hence inevitably suffered from running instability due to environmental perturbation on polarization states \cite{Huang2012}. Additionally, the initiation of synchronization operation usually required careful tuning of evolving polarization states inside the laser resonator \cite{Tsai2013, Sotor2014}. 

In this letter, we devised an all-polarization-maintaining synchronization system for mode-locked fiber lasers, which featured self-starting operation and long-term stability. The synchronized Er- and Yb-doped fiber lasers was constructed based on the nonlinear amplifying loop mirrors. Additionally, the Sagnac-type loops of the two laser cavities were jointed by a common section of single-mode fiber. The co-propagating pulses would induce non-reciprocal phase shift difference that facilitated the passive synchronization between the two dual-color fiber lasers. Consequently, a record-long tolerance of cavity-length mismatch up to 16.2 mm was achieved, which was at least tenfold larger than previously demonstrated values in the case of comparable pulse durations. Thanks to the high-speed nonlinear response, tight timing locking was manifested with a small relative timing jitter of 31 fs within a Fourier frequency range from 0.1 Hz to 1 MHz. The compact and robust synchronization system of two-color mode-locked fiber lasers would constitute practical light sources for expanding applications outside of a laboratory setting.

\begin{figure}[t!]
\centering
\includegraphics[width=0.95\columnwidth]{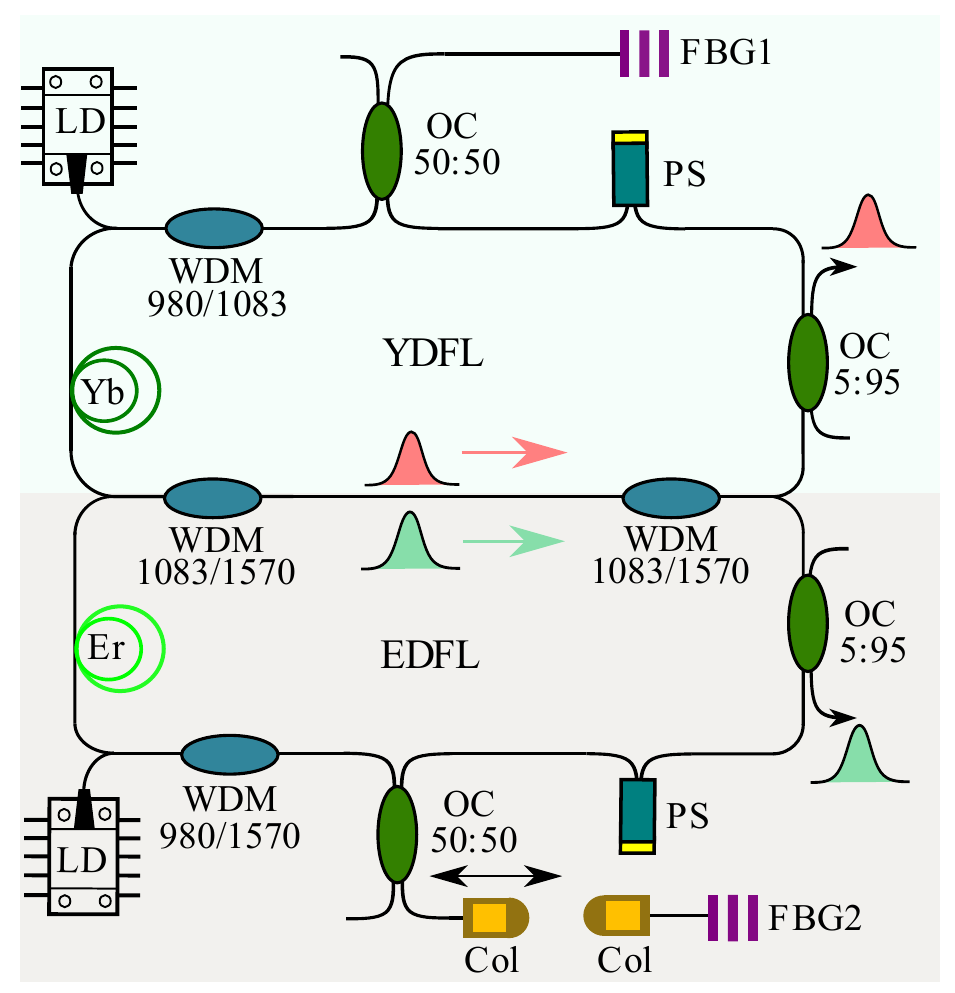}
\caption{Experimental schematic for all-polarization-maintaining synchronization system. The Yb- and Er-doped fiber lasers can be independently mode-locked based on the nonlinear amplifying loop mirror. The Sagnac loops within the laser cavities are combined to share a common section of single-mode fiber. The resulting co-circulating dual-color pulses would induce additional non-reciprocal phase shift, which could result in passive timing synchronization with self-starting and long-term stable performance. LD: Laser diode; WDM: wavelength division multiplexer;  Yb: ytterbium-doped gain fiber;  Er: erbium-doped gain fiber; OC: output coupler; FBG: fiber Bragg grating; PS: phase shifter; Col: collimator.}
\label{fig1}
\end{figure}

Figure \ref{fig1} illustrates the schematic for experimentally implementing the polarization-maintaining (PM) synchronization system, which was comprised of an Yb-doped fiber laser (YDFL) and an Er-doped fiber laser (EDFL) in a shared-cavity configuration. Individually, the two fiber lasers could be passively mode-locked based on nonlinear amplifying loop mirror (NALM) inside the laser cavity. The essential mechanism relies on the intensity-dependent transmission of the Sagnac interferometer, which could be regarded as an equivalent saturable absorber \cite{Hansel2017}. In the Sagnac loop, the phase shift difference between two counter-circulating fields should accumulate to an integer multiple of 2$\pi$ for launching the mode-locked operation \cite{Hansel2017}. To this end, the gain fiber was placed asymmetrically relative to the 2$\times$2 balanced optical coupler as shown in Fig. \ref{fig1}. Additionally, a compact $\pi$/2 phase shifter (PS) was used to provide a predefined linear phase difference, which not only substantially reduced the mode-locking threshold, but also facilitated the self-initiation of mode-locked operation \cite{Jiang2016}. In contrast to NPR-based mechanism, the NALM-based mode-locking favors all-polarization-maintaining implementation, which could offer better immunity to ambient disturbances.

\begin{figure}[b!]
\centering
\includegraphics[width=0.95\columnwidth]{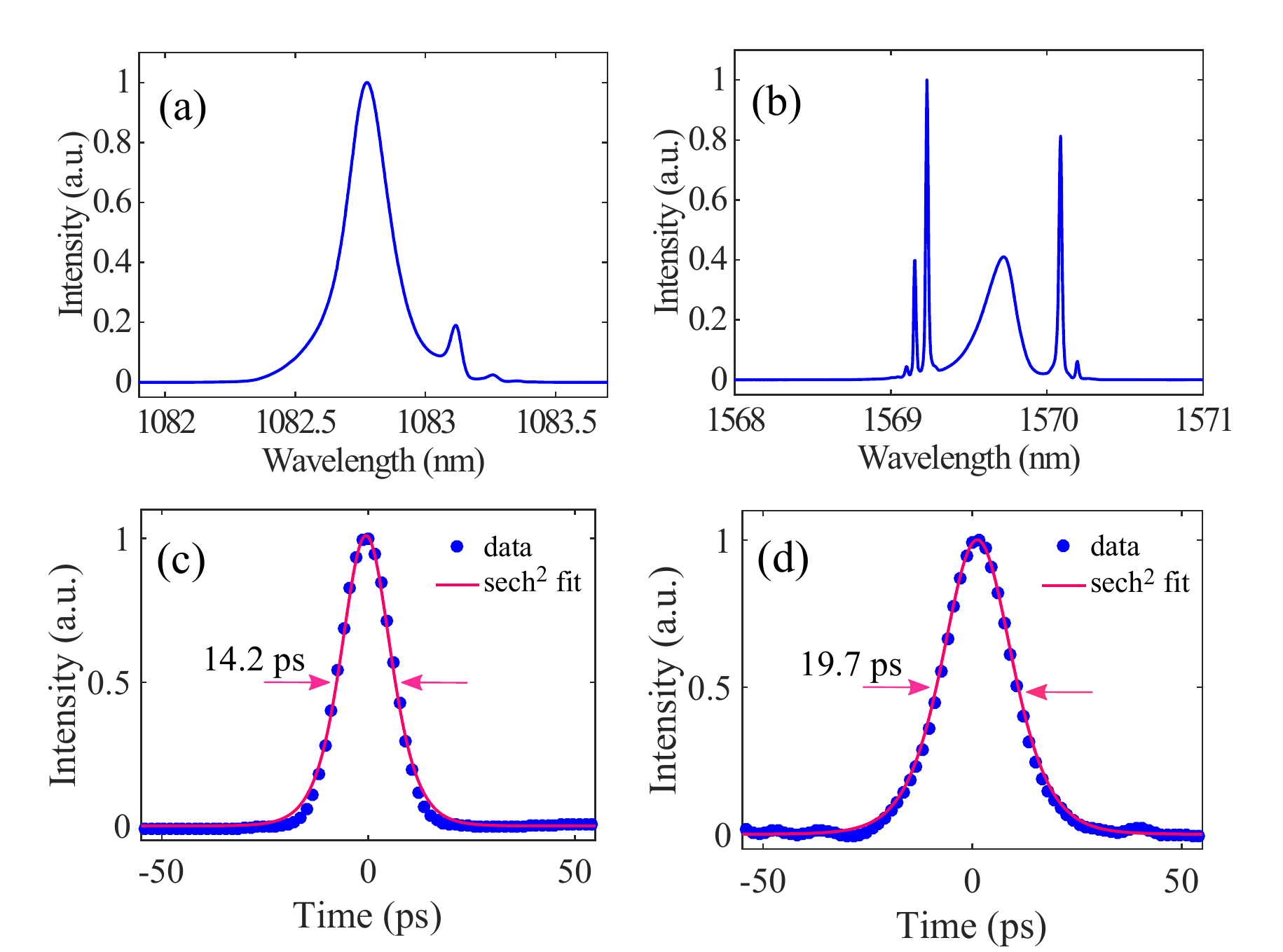}
\caption{ Experimental characterization of output pulses from YDFL (a, c) and EDFL (b, d) in the case of free-running mode-locked operation, including the measured optical spectra (a, b) and  corresponding auto-correlation traces (c, d).}
\label{fig2}
\end{figure}

We first examine pulse characteristics for the two fiber lasers at the free-running mode. The self-starting mode-locking could be obtained at a pump power of 170 mW and 500 mW for the YDFL and EDFL, respectively. Inside the laser cavity, another optical coupler was used to extract the circulating pulses with a tapping ratio of 5\%. The output spectrum was restricted within the reflection window of the intra-cavity fiber Bragg grating (FBG1 or FBG2) with a 1-nm bandwidth. The optical spectrum for the YDFL was thus centered at 1082.8 nm with a full width at half maximum (FWHM) of 0.2 nm while the spectrum for the EDFL shown a central wavelength of 1569.7 nm and a FWHM of 0.3 nm, which are given in Figs. \ref{fig2}(a) and \ref{fig2}(b), respectively. The sharp peaks exhibited in the spectrum of the EDFL were characteristic Kelly sidebands for soliton pulses typically formed in a cavity with an anomalous net dispersion \cite{Huang2018}. In the temporal domain, the auto-correlation traces were measured as shown in Figs. \ref{fig2}(c) and \ref{fig2}(d). The pulse durations from the Yb- and Er-fiber lasers were inferred to be 9.2 ps and 12.8 ps after taking into account the scaling factor of 1.54 under the assumption of sech$^2$ intensity envelop.

To synchronize the dual-color pulses from two mode-locked fiber lasers, the cavity-length mismatch, corresponding to the relative repetition rate, was imperative to be engineered in an appropriate range. As depicted in Fig. \ref{fig1}, a pair of collimators were mounted on a translation stage with a precision of 10 $\mu$m. The repetition rates were accurately recorded by using a spectral analyzer (Agilent N9000A) and thus set around 17.3 MHz.

\begin{figure}[t!]
\centering
\includegraphics[width=1\columnwidth]{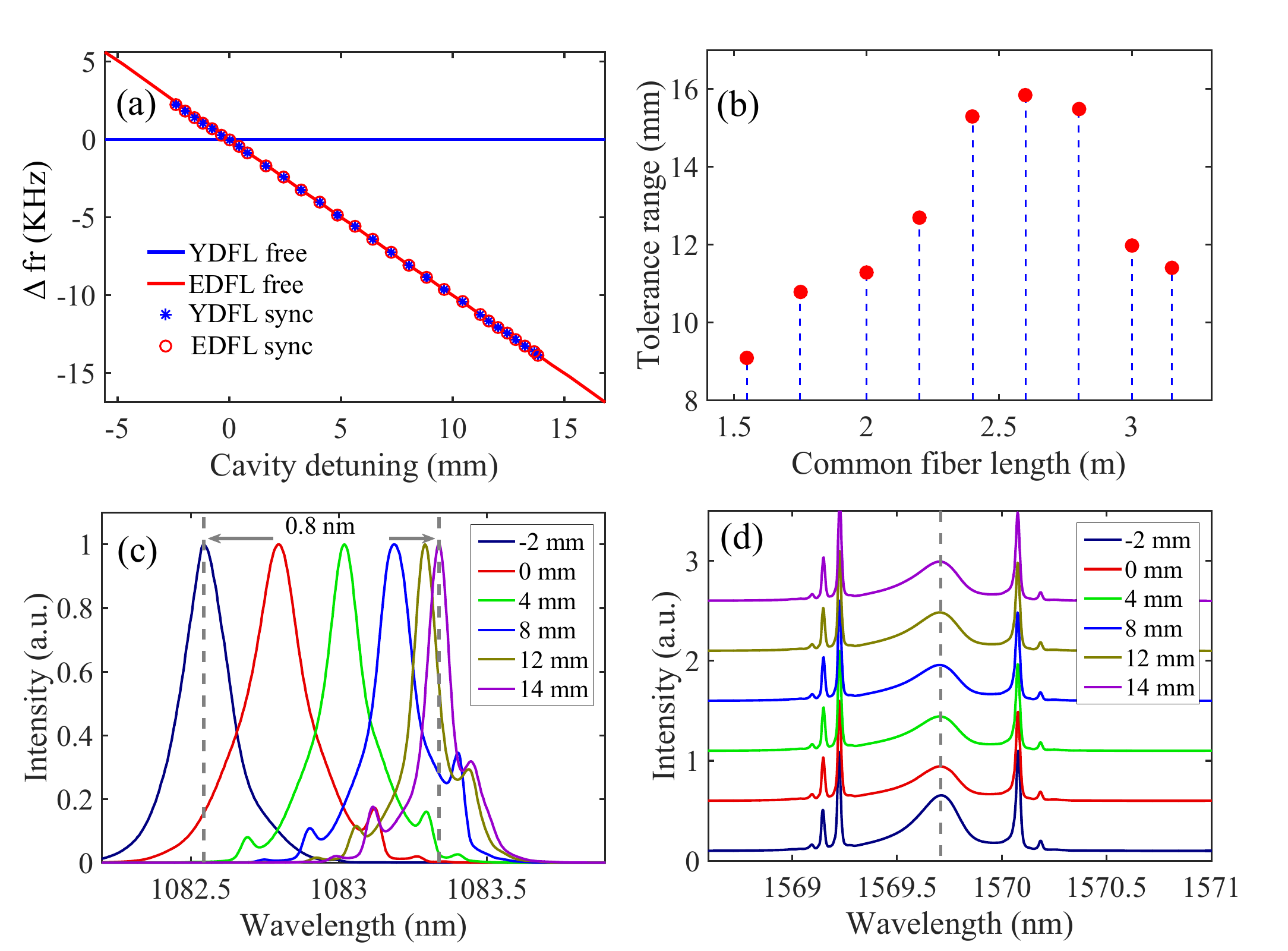}
\caption{(a) Measured repetition rates for Er and Yb fiber lasers as a function of the cavity-length variation of EDFL. The solid lines indicate the behavior for two respectively mode-locked laser without synchronization. The asterisk and circle symbols correspond to simultaneously mode-locked operation with a tight timing synchronization. Note that the repetition rates are shifted by the value of free-operation YDFL at 17.3 MHz. (b) Tolerance range of cavity-length mismatch varies with the common-fiber length. Optical spectra of YDFL (c) and EDFL (d) with various cavity-length detuning of EDFL.}
\label{fig3}
\end{figure}

Now the dual-color fiber laser system is ready for synchronization. In the experiment, the self-starting mode-locked YDFL was used to provide triggers in a digital oscilloscope. Then we monitor the output pulses from the EDFL while simply increasing the pump power of the laser diode. Temporal synchronization could be identified once the signal pulse waveforms were clearly displayed without swinging on the oscilloscope. Furthermore, the repetition rates for Er and Yb lasers were measured as varying the cavity length of the EDFL. The tolerance range shown in Fig. \ref{fig3}(a) unprecedentedly reached to 16.2 mm, which was more than one order of magnitude longer than previously demonstrated values for comparable pulse durations \cite{Huang2012, Zhang2011, Sotor2014}. In the conventional configurations based on either NPR \cite{Huang2012} or SA \cite{Sotor2014}, the intra-cavity polarization evolution would be inevitably altered by the cavity-length variation, which might terminate the synchronous mode locking. In contrast, this issue could be effectively mitigated by using the PM mode-locking mechanism based on NALM. In this case, the induced phase shift difference could almost be invariant for the dual-color co-propagating pulses, which ensured the stable passive synchronization for a much larger cavity-length detuning. Additionally, the PM-based structure offered enhanced immunity to the ambient polarization perturbation, thus leading to a robust and tight synchronization for days. Table \ref{table1} summarizes recent works on the passive synchronization to deliver a direct comparison in terms of various crucial figures of merit.

\begin{table*}[t!]
\centering
\caption{Comparison of passive synchronization performance for mode-locked fiber lasers with comparable pulse durations. SWNT: Single-wall carbon nanotube; GSA: Graphene saturable absorber; SESAM: Semiconductor saturable absorber mirror.}
\centering
\label{table1}
\begin{tabular}{p{0.08\linewidth}p{0.09\linewidth}p{0.12\linewidth}p{0.08\linewidth}p{0.14\linewidth}p{0.12\linewidth}p{0.11\linewidth}p{0.05\linewidth}}
\hline
Ref. & Wavelength (nm) & Pulse duration (ps) & Rep. rate (MHz) & Timing jitter (fs)/ Bandwidth (KHz) & Tolerance range (mm) & Mode-locking mechanism & Fiber type\\
\hline
This work & 1083/1570 & 9.2/12.8 & 17.3 & 31/1000 & 16 & NALM & PM\\
\cite{Zhang2011} & 1067/1535 & 6.1/2.1 & 13.1 & 600/4 & 1.4 & SWNT & NPM\\
\cite{Huang2018} & 1064/1570 & 19.5/12.1 & 20.3 & 26/1000 & 0.8 & NALM & PM\\
\cite{Sotor2014} & 1559/1938 & 0.9/1.6 & 20.5 & 67/10 & 0.78 & GSA & NPM\\
\cite{Huang2012} & 1040/1564 & 5.6/8.8 & 17.6 & 45/100 & 0.05 & NPR & NPM\\
\cite{Rusu2004} & 1040/1550 & 13/0.2 & 29.0 & $\backslash$ & 0.02 & SESAM & NPM\\
\hline
\end{tabular}
\end{table*}

As shown in Fig. \ref{fig3}(a), the repetition rate of the YDFL tightly followed that of the EDFL, thereby the Er-fiber laser acted as a master laser similarly in the master-slave arrangement \cite{Yoshitomi2006, Huang2012}. Due to different losses presented in the two fiber lasers, the output average power was measured to be 140 $\mu$W and 1.3 mW for YDFL and EDFL, respectively. Consequently, the pulse energy of the Er-fiber laser was about ninefold stronger than that of the Yb-fiber laser. The master feature of EDFL was also manifested in the accompanying change of optical spectra while varying the cavity length as given in Figs. \ref{fig3}(c) and (d). Specifically, the spectrum of EDFL remained immobile while the YDFL exhibited a 0.8-nm spectral shift within the capture range. It was the XPM-induced wavelength shift that compensated the cavity-length variations. The cavity round-trip time was thus self-adapted due to the group velocity dispersion \cite{Wei2002}. We note that the achieved tolerance range here was significantly longer than that in our previous demonstration \cite{Huang2018}. One possible reason might be the use of a larger-bandwidth FBG, thus supporting wider wavelength shift. Optimizing the FBG bandwidth could further augment the capture range albeit that too large filtering bandwidth might cause difficulty to achieve mode-locking \cite{Jiang2016}. Additionally, the shared-cavity configure facilitated a much compact layout because no additional amplifier was needed for the pulse injection as typically required in the maser-slave arrangement.

\begin{figure}[b!]
\centering
\includegraphics[width=0.8\columnwidth]{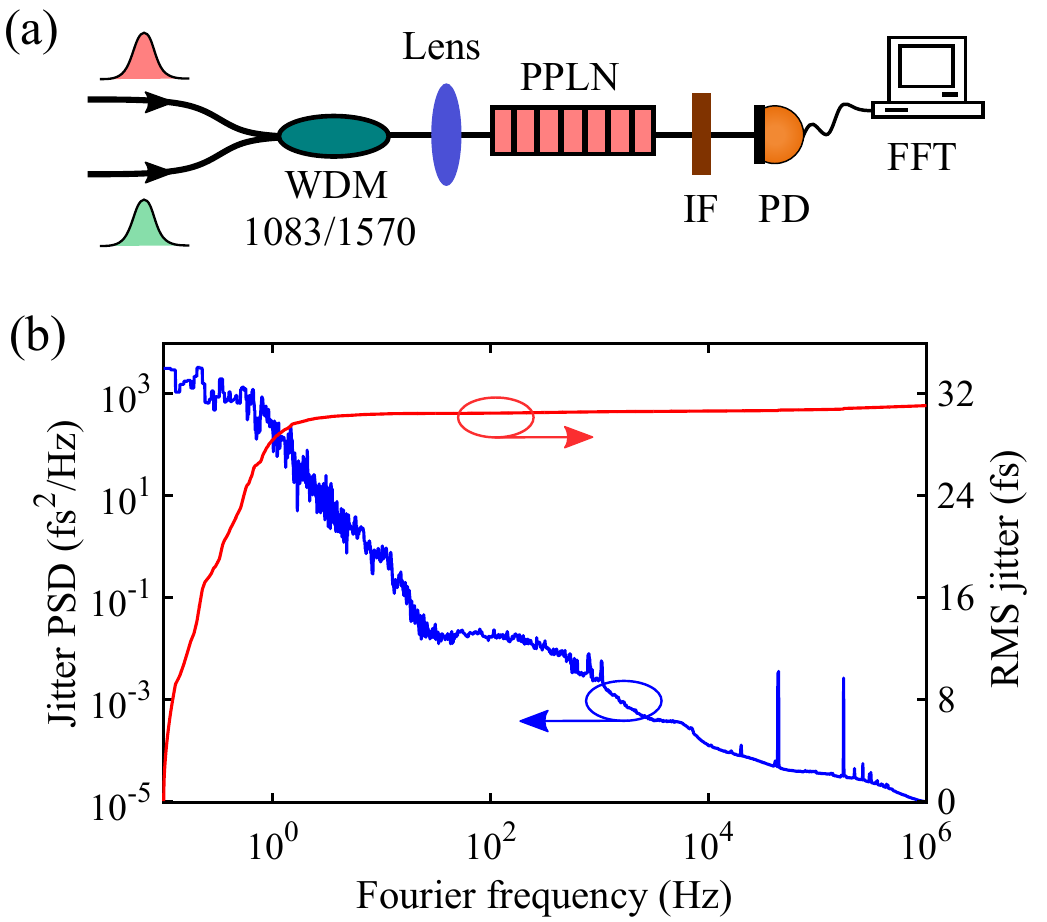}
\caption{(a) Characterization of relative timing jitter based on cross-correlation technique. PPLN: periodically-poled lithium niobate crystal; IF: interference filter; PD: photodiode; FFT: fast Fourier transform. (b) Power spectral density was recorded for the sum-frequency-generated signal at the half maximum of the correlation trace, which resulted in an integrated timing jitter of 31 fs within the Fourier domain from 0.1 Hz to 1 MHz.}
\label{fig4}
\end{figure}

In the following, we will investigate the dependance of cavity-mismatch tolerance on the geometric length of the single-mode fiber shared by two fiber lasers. As shown in Fig. \ref{fig1}, the two-color pulses were combined and divided by two wavelength division multiplexers (WDMs). The aforementioned tolerance range was obtained with a common fiber length of 2.6 m. It could be seen from Fig. \ref{fig3}(b) that degrading performance would be presented with a deviation from the optimal length. In the experiment, the group velocity mismatch between the pulses at 1083 nm and at 1570 nm was about 1.7 ps/m \cite{Huang2018}, corresponding to a walk-off length about 6 m for a 10-ps pulse duration. Therefore, the walk-off between the two-color pulses should be negligible within the investigated range from 1.55 to 3.15 m. Consequently, the nonlinear XPM-induced phases would be accumulated along the whole section of the shared fiber. Before reaching the turning point, the tolerance range would increase with the fiber length. Intuitively, the larger introduced nonlinear phase shift would lead to the enhancement of the pulling effect between two laser pulses, thus favoring a more robust passive synchronization. However, superabundant accumulated phases would break the stable mode-locking status, which is unique for the NALM-based fiber lasers. It was also intriguing to find that synchronized harmonic mode-locking could be obtained in the presence of a longer common fiber. In this case, the split pulses would possess a lower energy, thus rendering an appropriate nonlinear phase to initiate again the passive synchronization. 

Finally, the relative timing jitter between the synchronized pulses was characterized by using the optical cross-correlation technique. As shown in Fig. \ref{fig4}(a), the two-color pulses were spatially combined by a WDM before being focused onto a periodically-poled lithium niobate (PPLN) crystal for sum-frequency generation (SFG). The measurement of timing jitter was performed at the half amplitude of the cross-correlation trace, where the temporal variation would be linearly translated into the intensity fluctuations. Then the registered SFG signal was processed with Fourier transformation by a signal source analyzer (AnaPico APPH6040). The resulting power spectral density was presented in Fig. 4(b) with an average of three recorded traces. The integrated root-mean- square (RMS) timing jitter of 31$\pm$2 fs was obtained from 0.1 Hz to 1 MHz, which indicated tight synchronization enabled by our presented synchronization scheme. To go beyond the achieved timing jitter, possible improvement can resort to the adopt of shorter pulses and further suppress of the residual low-frequency noises. Therefore, hybrid synchronization scheme was envisioned to approach a sub-femtosecond jitter by combing the low-bandwidth active locking and the fast-response nonlinearity \cite{Tsai2013}.

To conclude, we have presented a novel scheme to realize a passively synchronized dual-color fiber laser system based on all-polarization-maintaining architecture. The implemented synchronization system exhibited desirable features such as compact layout, self-starting operation, low timing jitter, and long-term stability. In particular, the cavity-length-mismatch tolerance reached to an unprecedented range of 16.2 mm, which was at least tenfold larger than the previously reported values. Notably, a small relative timing jitter of 31 fs was obtained without any measures for temperature stabilization and vibration isolation. Therefore, the tight and robust synchronization fiber lasers would be attractive in applications beyond laboratorial operations. Moreover, the presented scheme is readily modified to access wavelengths around 2 $\mu$m with holmium- and thulium-doped fibers as the laser gain medium \cite{Liao2018}. By combining the techniques of nonlinear frequency mixing and intra-cavity wavelength tuning, multi-color ultrafast optical pulses could be engineered with a broadband spectral tunability. Such light sources would facilitate immediate application in coherent Raman scattering microscopy \cite{Nose2014}.

\section*{Funding} National Key Research and Development Program (2018YFB0407100), Program for Professor of Special Appointment (Eastern Scholar) at Shanghai Institutions of Higher Learning, Science and Technology Innovation Program of Basic Science Foundation of Shanghai (18JC1412000).

\end{document}